\documentclass[prd,aps,twocolumn,a4paper,floatfix,nofootinbib,preprintnumbers]{revtex4-1}

\usepackage[utf8]{inputenc}
\usepackage{graphicx,psfrag}
\usepackage{mathrsfs}
\usepackage{amsmath,amsfonts,amssymb}
\usepackage{multirow} 
\usepackage{diagbox}
\usepackage{comment}
\usepackage{xcolor}
\usepackage{enumerate}
\usepackage{booktabs}

\usepackage{hyperref}
\hypersetup{
    colorlinks = true,
    linkcolor = {blue},
    citecolor = {blue},
    urlcolor = {blue},
    linkbordercolor = {white},
    }

\usepackage{color}
\definecolor{cyan}{rgb}{0,0.9,0.9}
\definecolor{orange}{rgb}{0.9,0.5,0}
\definecolor{magenta}{rgb}{1,0,1}
\definecolor{purple}{rgb}{0.8,0.4,0.8}
\definecolor{gray}{rgb}{0.8242,0.8242,0.8242}
\definecolor{mgreen}{rgb}{0.1,0.8,0.1}

\usepackage[normalem]{ulem}
\newcommand{\fmiq}{\,\mathrm{fm}^{-3}}

\begin{document}
\preprint{LA-UR-21-27560}

\title{Quantifying modeling uncertainties when combining multiple gravitational-wave detections from binary neutron star sources}

\author{Nina \surname{Kunert}$^{1}$}
\email{nkunert@uni-potsdam.de}
\author{Peter~T.~H. \surname{Pang}$^{2,3}$}
\author{Ingo \surname{Tews}$^{4}$}
\author{Michael W.~\surname{Coughlin}$^{5}$}
\author{Tim \surname{Dietrich}$^{1,6}$}

\affiliation{${}^1$Institute for Physics and Astronomy, University of Potsdam, D-14476 Potsdam, Germany}
\affiliation{${}^2$Nikhef, Science Park 105, 1098 XG Amsterdam, The Netherlands}
\affiliation{${}^3$Institute for Gravitational and Subatomic Physics (GRASP), Utrecht University, Princetonplein 1, 3584 CC Utrecht, The Netherlands}
\affiliation{${}^4$Theoretical Division, Los Alamos National Laboratory, Los Alamos, NM 87545, USA}
\affiliation{${}^5$School of Physics and Astronomy, University of Minnesota, Minneapolis, Minnesota 55455, USA}
\affiliation{${}^6$Max Planck Institute for Gravitational Physics (Albert Einstein Institute), Am M\"uhlenberg 1, Potsdam 14476, Germany}

\date{\today}

\begin{abstract}
With the increasing sensitivity of gravitational-wave detectors, we expect to observe multiple binary neutron-star systems through gravitational waves in the near future. 
The combined analysis of these gravitational-wave signals offers the possibility to constrain the neutron-star radius and the equation of state of dense nuclear matter with unprecedented accuracy.
However, it is crucial to ensure that uncertainties inherent in the gravitational-wave models will not lead to systematic biases when information from multiple detections are combined.
To quantify waveform systematics, we perform an extensive simulation campaign of binary neutron-star sources and analyse them with a set of four different waveform models. 
For our analysis with 38 simulations, we find that statistical uncertainties in the neutron-star radius decrease to $\pm 250\rm m$ ($2$\% at $90$\% credible interval) but that systematic differences between currently employed waveform models can be twice as large. 
Hence, it will be essential to ensure that systematic biases will not become dominant in inferences of the neutron-star equation of state when capitalizing on future developments.  
\end{abstract}

\maketitle

\section{Introduction}
\label{sec:intro}

Gravitational waves (GWs) emitted from binary neutron-star (BNS) coalescences allow us to probe the equation of state (EOS) of dense nuclear matter.
This was successfully demonstrated by the LIGO-Virgo Collaboration and other research groups following the first GW observation of a BNS system, GW170817, using Bayesian analyses of the GW signal~\cite{TheLIGOScientific:2017qsa,Abbott:2018wiz,Abbott:2018exr,LIGOScientific:2018mvr,De:2018uhw}. 
Such constraints have been further improved in numerous multi-messenger analyses, e.g., Refs.~\cite{Bauswein:2017vtn,Annala:2017llu,Most:2018hfd,Abbott:2018exr,Radice:2018ozg,Dai:2018dca,Hinderer:2018pei,Capano:2019eae, Dietrich:2020lps, Legred:2021hdx, Raaijmakers:2021uju, Huth:2021bsp}, by incorporating data from associated electromagnetic observations, AT2017gfo and GRB170817A~\cite{Monitor:2017mdv}, nuclear-physics computations~\cite{Hebeler:2013nza,Annala:2017llu,Tews:2018kmu}, nuclear-physics experiments~\cite{Danielewicz:2002pu,Russotto:2016ucm, PREXII}, as well as radio and X-ray observations of isolated neutron stars (NSs)~\cite{Antoniadis:2013pzd,NANOGrav:2017wvv,Fonseca:2021wxt,Miller:2019cac,Riley:2019yda,Miller:2021qha,Riley:2021pdl}.

Extracting information from observational data always requires certain modeling assumptions. 
For example, to infer information from the measured GW data, it is necessary to cross-correlate the observed GW signal with theoretical models describing the compact binary coalescence for various binary parameters.
Following this approach, the matching introduces systematic uncertainties that originate from the approximations made to describe the general relativistic two-body problem. 
These approximations range from an analytical description using the Post-Newtonian (PN) framework~\cite{Blanchet:2013haa}, the effective-one-body (EOB) model~\cite{Buonanno:1998gg, Buonanno:2000ef} to numerical-relativity simulations, e.g., Refs.~\cite{Baiotti:2016qnr,Dietrich:2020eud}. 
Since it is expected that statistical uncertainties will be reduced for high signal-to-noise ratio (SNR) signals or when multiple signals are combined, systematic uncertainties introduced by the waveform models will become increasingly prominent and it is crucial to understand all sources of systematic uncertainties for a reliably quantification of EOS constraints.

Previous studies have shown that EOS constraints based on tidal deformabilities extracted from GW170817 were dominated by statistical uncertainties, e.g., Ref.~\cite{Abbott:2018exr}, and that systematic biases were under control, i.e., noticeably smaller than statistical ones.
For example, Ref.~\cite{Dudi:2018jzn} performed an injection study to investigate systematic uncertainties from different GW models and found systematic biases for GW170817 to be small. However, for similar sources observed at Advanced LIGO and Advanced Virgo design sensitivity, different waveform models can lead to noticeable biases, i.e., the recovered 90\% credible intervals would not contain the injected values.
Similarly, Ref.~\cite{Samajdar:2018dcx} used simulated, non-spinning GW170817-like sources measured at design sensitivity and found that for unequal masses the obtained tidal parameters can get noticeably biased.
This work was extended by Ref.~\cite{Samajdar:2019ulq} by studying the imprint of precession and source localization obtained through a possible electromagnetic counterpart.
Ref.~\cite{Gamba:2020wgg} found that existing GW waveform models used for the analysis of GW170817 will be dominated by systematic uncertainties for SNRs above $80$.  
Finally, very recently, Ref.~\cite{Chatziioannou:2021tdi} discussed numerous systematic biases that enter GW analyses outlining the importance of waveform systematics. 

As pointed out in, e.g., Refs.~\cite{DelPozzo:2013ala, Agathos:2015uaa, Lackey:2014fwa, Wysocki:2020myz}, even low SNR signals can be used and combined to constrain the tidal deformability parameter to an accuracy of $\sim$ 10\% with only a few tens of detections; cf.~also Refs.~\cite{Favata:2013rwa, Wade:2014vqa}. Using such a procedure, systematic biases are introduced through ``stacking,'' i.e., combining multiple GW measurements. 
However, to our knowledge no study to date has addressed waveform systematics introduced through the stacking of signals within a realistic injection study. 
Here, we address this issue and determine at which point systematic biases dominate.
We use a set of $38$ simulated signals analysed with four different waveform models and perform a total of $152$ BNS parameter estimation simulations assuming Advanced GW detectors at design sensitivity \cite{TheLIGOScientific:2014jea, TheVirgo:2014hva}. 
Throughout this work, geometric units are used by setting $G = c = 1$. Further notations are $M = M_A + M_B$ for a system's total mass, $q = M_A/M_B$ for the mass ratio, and $\Lambda_A$, $\Lambda_B$ for individual tidal deformabilities of the stars in a binary.

\section{Methods}
\label{sec:methods}

\paragraph*{\textbf{Combining information from multiple detections.}}
Extracting tidal effects from GW data requires information about the EOS. 
In this study, we will use EOSs that are constrained by chiral effective field theory (EFT) at low densities~\cite{Tews:2018kmu, Tews:2018iwm}. 
Chiral EFT is a systematic theory for nuclear forces that provides an order-by-order scheme for the interactions among neutrons and protons~\cite{Epelbaum:2008ga, Machleidt:2011zz}. 
These interactions can then be used in microscopic studies of dense matter up to densities of $\sim 2$ times the nuclear saturation density ($n_{\rm sat}=0.16\fmiq$). 
In Ref.~\cite{Dietrich:2020lps}, $5000$ EOSs constrained by quantum Monte Carlo calculations using chiral EFT interactions up to $1.5 n_{\rm sat}$ were computed. 
For this article, we employ the most-likely EOS of Ref.~\cite{Dietrich:2020lps} for all of our BNS injections.
During the parameter estimation, instead of sampling masses and tidal deformabilities independently, we sample from the same set of $5000$ EOSs. 
These EOSs relate masses and tidal deformabilities based on nuclear-physics information. 
The tidal deformabilities are then computed for a given mass and EOS via
\begin{equation}
    p(\Lambda_i | m_i, {\rm EOS}) = \delta(\Lambda_i - \Lambda(m_i; {\rm EOS})),
\end{equation}
with $m_i$ and $\Lambda_i$ denoting the mass and tidal deformability of the stars.
In addition to the tidal behavior, the EOS also determines the maximum allowed mass $M_{\rm max}$ for NSs. 
Therefore, we choose a uniform NS mass distribution given by
\begin{equation}
    p(m_1, m_2|{\rm EOS}) = \frac{2\Theta(m_1 - m_2)\Theta(M_{\rm max} - m_1)}{(M_{\rm max} - M_{\rm min})^2},
\end{equation}
where $M_{\rm min}$ is the minimum NS mass and $\Theta$ is the Heaviside step function. We choose $M_{\rm min}$ to be 0.5$M_{\odot}$.

For the EOS prior probability, the mass measurements of PSR~J0348+0432~\cite{Antoniadis:2013pzd}, PSR~J1614-2230~\cite{Arzoumanian:2017puf}, and PSR~J0740+6620~\cite{Fonseca2021} are taken into account, similar to the approach outlined in Ref.~\cite{Dietrich:2020lps}. 
The prior gives rise to $R_{1.4}=12.24^{+1.21}_{-1.44}{\rm km}$ (at 90\% credibility). 
In contrast to \cite{Dietrich:2020lps}, the NICER observation of PSR~J0030+0451~\cite{Miller:2019cac,Riley:2019yda} and the upper bound on $M_{\rm max}$ derived from GW170817~\cite{Rezzolla:2017aly} are not included here to avoid masking the systematic uncertainties in the GW analysis by additional information.

Since the EOS is a common parameter, we can combine the information from multiple simulations to compute the combined posterior, $p_c(\vec{\theta})$. With $N_{\rm obs}$ detections $\{d_i\}$ it is given as
\begin{equation}\label{comb_posterior}
    p_c (\mathrm{EOS}| \{\vec{d}_i\}) = p (\mathrm{EOS})^{1-N_{\rm obs}} \prod_{i=1}^{N_{\rm obs}} p_i (\mathrm{EOS} | \{\vec{d}_i\}),
\end{equation}
in which $p_i(\mathrm{EOS}|\{d_i\})$ is the EOS posterior given the $i$-th simulation and $p(\mathrm{EOS})$ refers to the prior employed. 
To correct the selection bias introduced by non-uniform detectability across sources, we follow Ref.~\cite{Mandel:2018mve} to compute the joint posterior for the EOS as
\begin{equation}
    p_c (\mathrm{EOS}| \{\vec{d}_i\}) = p (\mathrm{EOS})^{1-N_{\rm obs}} \prod_{i=1}^{N_{\rm obs}} \frac{p_i (\mathrm{EOS} | \{\vec{d}_i\})}{\int d\vec{\theta} p_{\rm det}(\vec{\theta}) p(\vec{\theta}|{\rm EOS})},
\end{equation}
where $p_{\mathrm{det}}(\vec{\theta})$ is the probability of detecting the GW signal corresponding to the source parameters $\vec{\theta}$. 
In this work, a threshold SNR of $7$ is enforced for the detections and we estimate $p_{\rm det}(\vec{\theta})$ using a neural-network classifier as described in Ref.~\cite{Gerosa:2020pgy} trained on the BNS parameter distributions.

\paragraph*{\textbf{Waveform Models.}}
The frequency-domain representation of a gravitational waveform can be written as 
\begin{equation}
\tilde{h}(f)=\tilde{A}(f)e^{-{\rm i}\psi(f)}, \label{eq:h_FD}
\end{equation}
with the frequency $f$, the amplitude $\tilde{A}(f)$, and GW phase $\psi(f)$. The phase can be further decomposed into 
\begin{equation}
\label{eq:psi_omg}
 \psi(f) = \psi_{\rm pp} (f) + \psi_{\rm SO}(f) + \psi_{\rm SS}(f) + \psi_{\rm T}(f)+ \cdots \,, 
\end{equation}
with $\psi_{\rm pp}$ being the non-spinning point-particle
contribution, $\psi_{\rm SO}$ corresponding to contributions caused by spin-orbit coupling, $\psi_{\rm SS}$ to contributions caused by spin-spin effects, and $\psi_{\rm T}$ denoting the tidal effects present in the GW phase. We note that higher-order effects, e.g., cubic-in-spin, could also be included.

The dominant quantity describing EOS-related tidal effects on the GW signal is the mass-weighted tidal deformability \cite{Hinderer:2009ca, Hinderer:2007mb, Favata:2013rwa}
\begin{equation}
\label{eq:kappa}
\tilde{\Lambda} = \frac{32}{39} \left[\left(1+12\frac{X_B}{X_A}\right)\left(\frac{X_A}{C_A}\right)^5 k^A_2 + (A \leftrightarrow B) \right]  \ ,
\end{equation}
with the compactness parameters of the individual undisturbed stars $C_{A,B} = M_{A,B}/R_{A,B}$,
the Love numbers $k^{A,B}_2$~\cite{Hinderer:2007mb,Damour:2009vw,Binnington:2009bb}, and $X_{A,B}=M_{A,B}/(M_A+M_B)$. 
The leading-order PN contribution to the tidal phase, proportional to $\tilde{\Lambda}$, starts at the 5PN order, i.e., becomes most relevant at the late inspiral; cf.\ Fig.~2 of Ref.~\cite{Dietrich:2020eud}.
To quantify the modelling systematics, we employ four different GW waveform models: TaylorF2 (TF2), IMRPhenomD\_NRTidal (PhenDNRT), IMRPhenomD\_NRTidalv2 (PhenDNRTv2), SEOBNRv4\_ROM\_NRTidalv2 (SEOBNRTv2). 
These models are based on different point-particle and tidal descriptions and, hence, lead to different estimates on the intrinsic source properties. 
The variety of GW  models allows us to disentangle the effects of the tidal contribution, comparing PhenDNRT vs.\ PhenDNRTv2, and the point-particle information, comparing PhenDNRTv2 vs.\ SEOBNRTv2. 
The comparison with TF2 serves as a ``worst case'' scenario since point-particle and tidal contributions are varied with respect to the injected waveform set (see Appendix \ref{B}).

An easy interpretation of the tidal contribution for our employed models can be extracted from Figs.~3 and 4 of Ref.~\cite{Dietrich:2019kaq}. 
NRTidalv2 predicts larger tidal effects than TaylorF2 and smaller tidal effects than NRTidal for the same tidal deformability. 
Hence, it is expected that NRTidal models will predict smaller NS radii and TaylorF2 will potentially predict larger radii with respect to our reference model NRTidalv2. 
Considering the differences in the employed point-particle contributions, we refer to Fig.~3 of Ref.~\cite{Samajdar:2018dcx}, where it was found that when using the same tidal contribution, both TaylorF2 and SEOBNRv4\_ROM lead to smaller estimated tidal deformabilities than IMRPhenomD. 
A summary of the expected biases and the final results is given in Tab.~\ref{tab:expectation}.  

\begin{figure}
    \centering
    \includegraphics[width=1.\columnwidth]{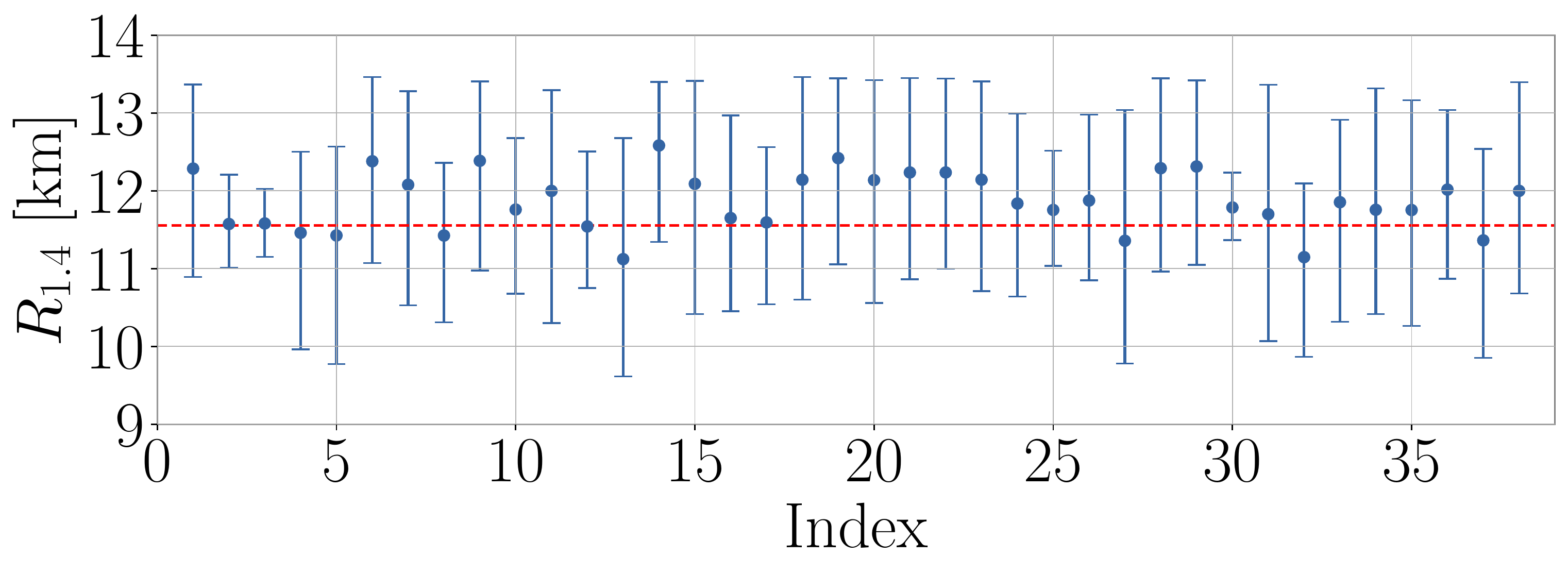}
    \caption{Constraints on the radius of a typical NS, $R_{1.4}$, from simulated raw data for the injection model PhenDNRTv2. 
    The radius for the injected EOS is shown as red dashed line.}
    \label{fig:raw_data_radius_PhenomDNRTidalv2}
\end{figure}

\paragraph*{\textbf{Injection setup.}}
We simulate a network of interferometers consisting of Advanced LIGO and Advanced Virgo at design sensitivity~\cite{TheLIGOScientific:2014jea, TheVirgo:2014hva}. 
The BNS sources in our injection set are uniformly distributed in a co-moving volume with the optimal network SNR $\rho \in [7, 100]$. 
The sources' sky locations $(\alpha, \delta)$ and orientations $(\iota, \psi)$ are placed uniformly on a sphere. 
Based on the observed BNS population, the spins of NSs are expected to be small~\cite{OShaughnessy:2006uzj}. 
We restrict the spin magnitudes of the two stars $(a_1, a_2)$ to be uniformly distributed, $a_i \in [-0.05, 0.05]$. 
The component masses are sampled from a uniform distribution of $m_{1,2} \in [1,2]~M_{\odot}$. 
For our $38$ BNS setups, we have used PhenDNRTv2 as injection model and employed the most-likely EOS of Ref.~\cite{Dietrich:2020lps}, leading to an injected radius of a typical $1.4M_\odot$ NS, $R_{1.4}$, of $\sim 11.55 \rm km$, corresponding to a dimensionless tidal deformability of $\Lambda_{1.4} \sim 292.5$.
We have used the four different models for the recovery, leading to a total of $152$ inference runs.

\section{Results}\label{sec:results}

\begin{figure}
    \centering
    \includegraphics[width=1.\columnwidth]{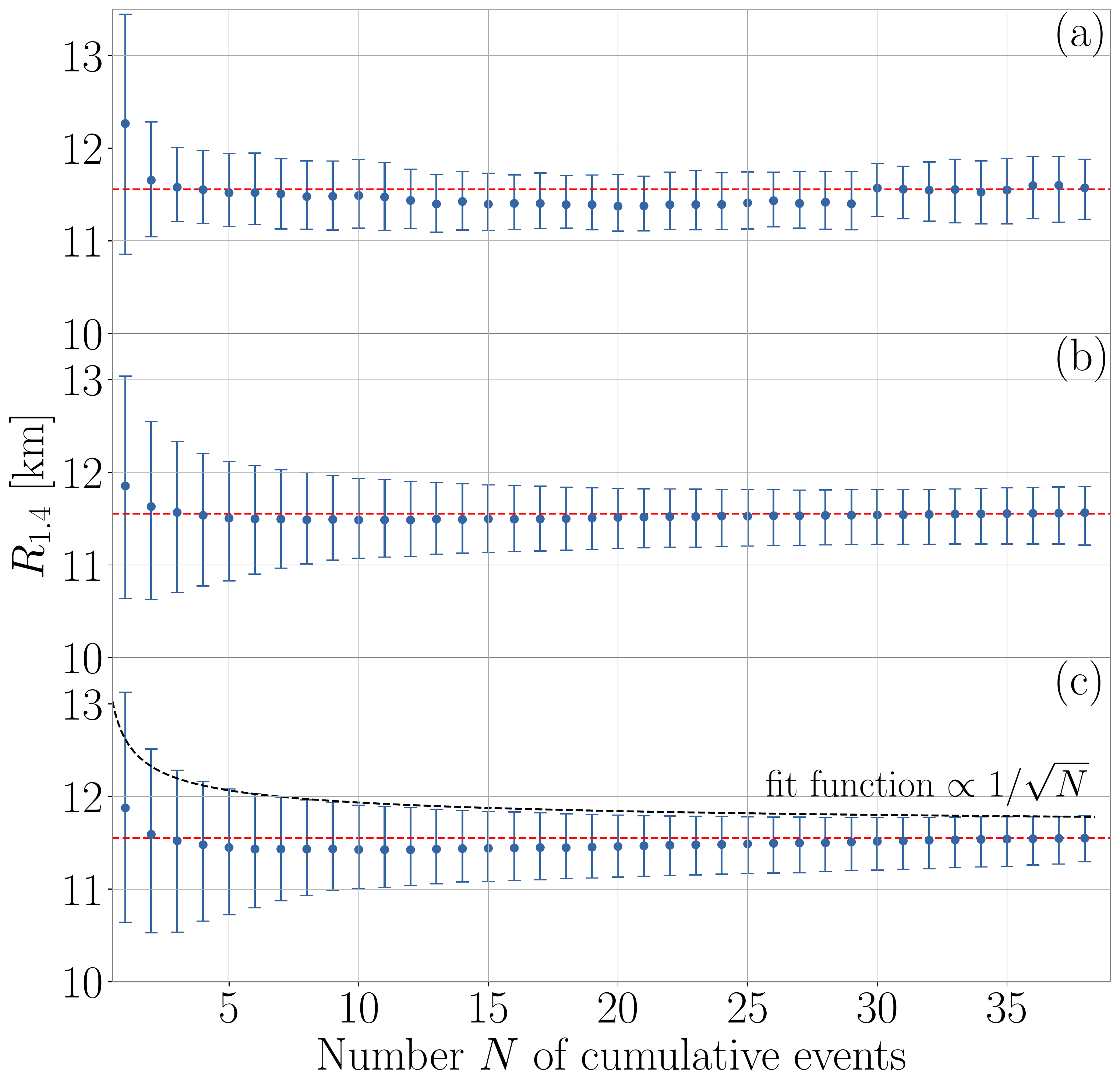}
    \caption{Overview of applied corrections for injection model PhenDNRTv2: (a) combined injection data, (b) combined and randomly reshuffled injection data, and (c) same data as before but corrected for selection effects as detailed in Sec.~\ref{sec:methods}. 
    The injected NS radius is 11.55~km (red-dashed line).  
    }
    \label{fig:radius_PhenomDNRTidalv2}
\end{figure}

\paragraph*{\textbf{Radius measurements and intrinsic biases.}}
In Fig.~\ref{fig:raw_data_radius_PhenomDNRTidalv2}, we present our preliminary results for the NS radius, $R_{1.4}$, for the injection model PhenDNRTv2 for each individual injected GW~event denoted through a random identifier (index). 
The uncertainties reflect the 90\% confidence intervals. 
Depending on the source properties, the SNR and the particular noise realisation, different GW~events place tighter or weaker constraints on the neutron star radius. 
To improve our radius estimate on the underlying EOSs, one can combine multiple GW~events. 
In Fig.~\ref{fig:radius_PhenomDNRTidalv2}, we show the $R_{1.4}$~results for successively combined GW~events. 
Fig.~\ref{fig:radius_PhenomDNRTidalv2}(a) clearly shows that the combination of multiple GW events significantly reduces the uncertainty of a NS radius measurement. 
Furthermore, in order to avoid an arbitrary ordering effect in our simulated BNS population, we randomly shuffle the order (indexed events as shown in Fig.~\ref{fig:raw_data_radius_PhenomDNRTidalv2}) of the $38$ simulated events for $1000$ times and compute the median over all permutations; cf.~Fig.~\ref{fig:radius_PhenomDNRTidalv2}(b).

\begin{figure*}[t]
    \centering
    \includegraphics[width=\textwidth]{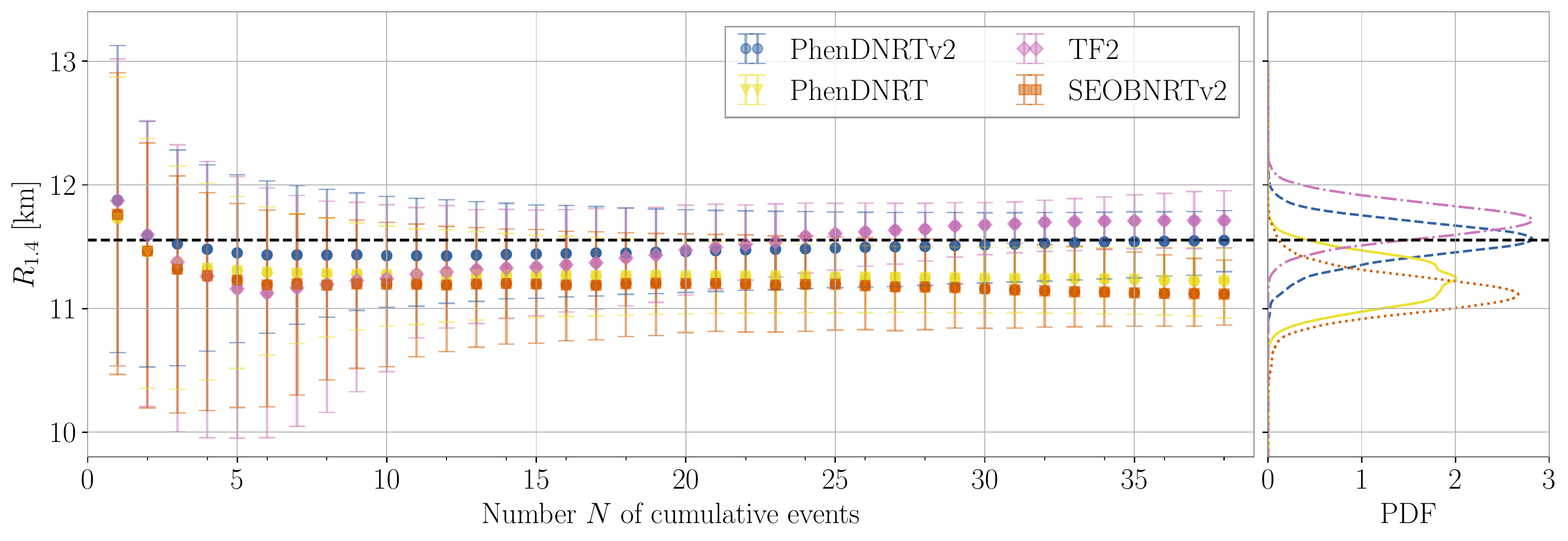}
    \caption{Left: radius constraints of a typical NS with $1.4 M_{\odot}$, $R_{1.4}$, versus number of successively combined GW~signals for the GW~models PhenDNRT, TF2, SEOBNRTv2, and the injection model PhenDNRTv2. Right: posterior distribution function (PDF) for each GW~model when all GW~signals are combined. 
    The injected EOS value is shown as black-dashed line. 
    }
    \label{fig:radius_combined}
\end{figure*}

At this stage, our study still includes an additional selection bias in our BNS population due to our inability of detecting arbitrarily weak GW signals~\cite{Mandel:2018mve}. 
Therefore, we are systematically more sensitive to sources with higher SNRs. 
This means that more massive binaries are favoured due to their high SNRs\footnote{We note that although the extrinsic parameters affect the SNR, they do not contribute to an uneven detectability for given intrinsic parameters.}.
Because these systems also correspond to lower tidal deformabilities, this selection effect could lead to smaller $R_{1.4}$-predictions. 
When correcting for this selection bias using the neural network classifier of Ref.~\cite{Gerosa:2020pgy}, we obtain our final result for $R_{1.4}$ shown in Fig.~\ref{fig:radius_PhenomDNRTidalv2}(c). 
As can be seen in Fig.~\ref{fig:radius_PhenomDNRTidalv2}(b)-(c), we recover the expected $\sim 1/\sqrt{N}$ falloff of the statistical uncertainty of the radius measurement, where $N$ denotes the number of combined GW detections. 
When all GW~events are combined, we obtain a final NS~radius estimate for the injection model PhenDNRTv2 of $11.55^{+0.24}_{-0.25} \rm km$ which is in perfect agreement with the injected value of $11.55 \rm km$ (red dashed line). 
The injection set corrections illustrated in the panels of Fig.~\ref{fig:radius_PhenomDNRTidalv2} were likewise applied to the other waveform models used in this study. 
The final NS radius measurements for all models are listed in Tab.~\ref{tab:expectation} and are shown in Fig.~\ref{fig:radius_combined}. 
From our NS radius measurement for the injection model PhenDNRTv2, we confirm the findings of Ref.~\cite{DelPozzo:2013ala} that the tidal deformability can be measured with a statistical uncertainty of $\sim$~10\% after a few tens of detections. 

\begin{table}[t]
\setlength{\tabcolsep}{6pt}
\renewcommand{\arraystretch}{1.2}
 \caption{\label{tab:expectation} 
 Overview of expected and measured NS radius measurements, $R_{1.4}$, and systematic shifts, $\Delta R_{1.4}$, for our selected GW models with respect to PhenDNRTv2. $\circ$ marks no expected bias, $\Uparrow$ a larger expected radius estimate, $\Downarrow$ a smaller expected radius, and $\Updownarrow$ denotes cases in which competing effects are present. 
 The injected NS~radius is $11.55$~km. 
 Our results for $\Delta R_{1.4}$ in the last column are based on the shift of the median value from the injected value.}
 \begin{tabular}{l|c c c| c c} 
 Model & $\psi_{\rm pp}$ & $\psi_{\rm T}$ & $\psi$ & $R_{1.4}$ [km] & $\Delta R_{1.4}$ [m]\\ 
 \hline
 PhenDNRTv2 & $\circ$ & $\circ$ & $\circ$ & $11.55^{+0.24}_{-0.25}$ & $-$3 \\ 
 PhenDNRT   & $\circ$ & $\Downarrow$ & $\Downarrow$ & $11.22^{+0.26}_{-0.30}$ & $-$329\\ 
 SEOBNRTv2  & $\Downarrow$ & $\circ$ & $\Downarrow$ & $11.12^{+0.28}_{-0.25}$ & $-$437 \\ 
 TF2        & $\Downarrow$ & $\Uparrow$ & $\Updownarrow$ & $11.71^{+0.24}_{-0.26}$ & $+$158\\
 \hline
 \end{tabular}
\end{table}

\paragraph*{\textbf{Waveform-model systematics.}} In Tab.~\ref{tab:expectation}, we show our expectations of potential systematic biases originating from different modelling assumptions, which we compare with our results for all waveform models in Fig.~\ref{fig:radius_combined}. 
The injection model PhenDNRTv2 perfectly recovers the injected value, but all other models employed, introduce a considerable bias in the NS radius measurement. 
PhenDNRT predicts lower NS radii which can be explained by the different tidal descriptions in the PhenDNRT and PhenDNRTv2, leading to overall smaller tidal deformabilities, and, hence, to smaller NS radii. 
Due to this systematic bias and decreasing statistical uncertainties, the model recovers the injected value within the 90\% credible interval only when less than 20 GW~events are combined. 
Combining all 38 GW~events, PhenDNRT recovers a NS~radius of $11.22^{+0.26}_{-0.30} \rm km$ which is lower than the injected value by $\sim 300 \rm m$. 
We note that the bell-shaped PDF for PhenDNRT in Fig.~\ref{fig:radius_combined} is an effect of our selection bias correction and is not present when excluding this correction.
We find the same trend towards smaller NS radii for SEOBNRTv2. 
Here, the different point-particle phase description in the model predicts smaller tidal deformabilities and, hence, smaller NS radii. 
From the results in Fig.~\ref{fig:radius_combined}, we find that the model is able to recover the injected NS radius in the 90\% credible interval when less than 30 GW~events are combined. 
The SEOBNRTv2 model predicts a final NS radius of $11.12^{+0.28}_{-0.25} \rm km$ when all injections are combined, resulting in a systematic shift of $\sim 400 \rm m$ towards smaller NS radii. 
In comparison to the injection model PhenDNRTv2, the systematic biases present in PhenDNRT and SEOBNRTv2 lead to overall smaller NS radii when all GW~events are combined.
However, our analysis of the injection raw data shows that for some individual simulations PhenDNRT predicts slightly larger NS radii than PhenDNRTv2. 

Finally, TF2 shows an indefinite trend for $R_{1.4}$, giving smaller NS~radii for a smaller number of combined GW~events, whereas with more than $23$ combined GW~events TF2 overestimates the injected NS radius.
Notably, this model shows the largest uncertainties when less than 10 GW~events are combined. 
Using all 38~detections, TF2 predicts a NS~radius of $11.71^{+0.24}_{-0.26} \rm km$ which is larger than the injected value by $\sim 150 \rm m$. 
The trend of the TF2 model could be explained by the different point-particle and tidal-phase descriptions compared to the injection model PhenDNRTv2. 
Because in TF2 the point-particle sector is described up to 3.5PN order whereas tidal effects are included up to 7.5 PN order, the uncertainties of $\psi_{\mathrm{pp}}$ will dominate for smaller numbers of combined GW events, leading to smaller NS radii, while uncertainties of $\psi_{\mathrm{T}}$ at 7.5PN order begin to dominate when larger numbers of GW events are combined, leading to larger NS radii.
Therefore, the combined $R_{1.4}$-estimate for TF2 underestimates waveform systematics because competing effects balance out; see Tab.~\ref{tab:expectation} and Fig.~\ref{fig:radius_combined}.

Notably, the $R_{1.4}$-estimates when combining the first few events in Fig.~\ref{fig:radius_combined} follow the same pattern regardless which waveform model is employed. 
The first event results in an overestimation of $R_{1.4}$ because it is largely driven by our heavy-pulsar prior. 
The non-zero support of $\Lambda=0$ for low $\Lambda$ injections results in an underestimation of $R_{1.4}$ when including a few additional events. 

We find that our extracted systematic shifts of the NS radius of up to $\sim 400 \rm m$ are smaller than shifts $\sim \mathcal{O}(\mathrm{1km})$ estimated in previous studies~\cite{Gamba:2020wgg, Pratten:2021pro}. 
From a mock analysis of 15 sources, Ref.~\cite{Gamba:2020wgg} found that systematic errors of that order dominate over statistical errors for signals with SNR~$\gtrsim~80$ for current advanced detectors at design sensitivity. 
Ref.~\cite{Pratten:2021pro} found a similar result when neglecting dynamical tidal effects in their BNS population.
The differences with our findings could originate from the fact that in our parameter estimation runs we directly sample over an EOS set restricted by nuclear physics to obtain combined $R_{1.4}$-estimates.
Refs.~\cite{Gamba:2020wgg, Pratten:2021pro}, instead, used methods such as spectral parametrizations and universal relations for the EOS to translate binary tidal parameters into NS EOS and radius information.
Consequently, our parameter estimation accounts for more physical information on the NS radius which affects systematic biases. 
Moreover, systematics can become more pronounced once information across several events are combined. 
Appendix \ref{app:ppplot} shows that we do not find significant systematic biases when estimating mass and spin parameters with our waveform models.

\section{Conclusions}\label{sec:conclusion}

In this work, we performed a large injection campaign with a total of 152 BNS parameter estimation simulations to understand how the combination of information from multiple BNS detections will decrease statistical measurement uncertainties in the NS radius and to quantify the impact of waveform model systematics. 
For this purpose, we used four different waveform models with different point-particle and tidal phase descriptions. 
Based on 152 BNS simulations, our main findings are summarized below:\\
(i) We verified that both the combination of multiple GW sources and GW detections with high SNR will be influenced by the different modelling assumptions of existing GW models. \\
(ii) From a total number of 38 combined simulations, one might be able to constrain the NS radius with an accuracy of $\pm 250 \rm m$ for our injection model. \\
(iii) Our results showed that systematic effects substantially affect the NS radius measurement. 
In this work, these are strongest for the models PhenDNRT and SEOBNRTv2 (up to $\sim 440 \rm m$). 
Hence, these models cannot recover the injected NS radius in their 90\% credible interval when more than 20 or 30 GW~events are combined, respectively.\\
Overall, with increasing GW detector sensitivity and the projected BNS merger rate of $10^{+52}_{-10}$ detections per year estimated by Ref.~\cite{KAGRA:2020npa} or $9$–$88$ forecasted by Ref.~\cite{Petrov:2021bqm} for O4, waveform model systematics will influence the extraction of NS properties.
Moreover, it might be possible that systematic uncertainties in GW modelling could lead to inconsistencies of the measured NS properties between different messengers, in particular, when information from electromagnetic counterparts of potential multi-messenger observations or tighter constraints from future X-ray observations similar to Ref.~\cite{Miller:2019cac, Miller:2021qha} are included.
In fact, given the expectations of $1-13$ well-localized ($\leq 100 \rm \deg^2$ at $90$\% credible areas) GW detections for BNS systems with corresponding kilonova detection rates of $0.5-4.8$ events per year in O4 \cite{Petrov:2021bqm}, one can expect systematic effects when using currently available waveform models during the next observing runs.

\acknowledgements

We thank Tatsuya Narikawa and the LVK Extreme Matter group for fruitful discussions and comments on the study. PTHP is supported by the research programme of the Netherlands Organisation for Scientific Research (NWO). The work of I.T. was supported by the U.S. Department of Energy, Office of Science, Office of Nuclear Physics, under contract No.~DE-AC52-06NA25396, by the Laboratory Directed Research and Development program of Los Alamos National Laboratory under project numbers 20190617PRD1 and 20190021DR, and by the U.S. Department of Energy, Office of Science, Office of Advanced Scientific Computing Research, Scientific Discovery through Advanced Computing (SciDAC) NUCLEI program. Computational resources have been provided by the Los Alamos National Laboratory Institutional Computing Program, which is supported by the U.S. Department of Energy National Nuclear Security Administration under Contract No.~89233218CNA000001, and by the National Energy Research Scientific Computing Center (NERSC), which is supported by the U.S. Department of Energy, Office of Science, under contract No.~DE-AC02-05CH11231. We also acknowledge usage of computer time on Lise/Emmy of the North German Supercomputing Alliance (HLRN) [project bbp00049], on HAWK at the High-Performance Computing Center Stuttgart (HLRS) [project GWanalysis 44189], and on SuperMUC NG of the Leibniz Supercomputing Centre (LRZ) [project pn29ba].
M.~W.~Coughlin acknowledges support from the National Science Foundation with grant numbers PHY-2010970 and OAC-2117997. All posterior samples and results are available on \cite{Zenodo}.

\appendix

\section{Bayesian Inference}\label{A}

By using Bayes' theorem, the posterior $p(\vec{\theta} | d,\mathcal{H})$ on the parameters $\vec{\theta}$ under hypothesis $\mathcal{H}$ and with data $d$ is given by
\begin{equation}\label{Bayes_theorem}
        p(\vec{\theta} | d,\mathcal{H}) =  \frac{p(d|\vec{\theta},\mathcal{H})p(\vec{\theta}|\mathcal{H})}{p(d|\mathcal{H})},
\end{equation}
where $p(d|\vec{\theta},\mathcal{H})$, $p(\vec{\theta}|\mathcal{H})$, and $p(d|\mathcal{H})$ are the likelihood, prior, and evidence, respectively. 
The prior describes our knowledge of the parameters before the observation. It also naturally acts as a gateway for additional observations to be included as part of the analysis. 
The likelihood quantifies how well the hypothesis describes the data at a given point $\theta$ in the parameter space and the evidence marginalizes over the whole parameter space.
By assuming Gaussian noise, the likelihood $p(d|\vec{\theta},\mathcal{H})$ of a GW signal $h(\vec{\theta})$ with parameters $\vec{\theta}$ embedded in the data $d$ is given by~\cite{Veitch:2014wba}
\begin{equation}
        p(d|\vec{\theta},\mathcal{H}) \propto \exp\left(-2\int_{f_\textrm{low}}^{f_\textrm{high}}\frac{|\tilde{d}(f) - \tilde{h}(f;\vec{\theta})|^2}{S_n(f)}df\right),
\end{equation}
where $\tilde{n}(f)$ and $\tilde{h}(f)$ are the Fourier transform of $n(t)$ and $h(t;\vec{\theta})$, $S_n(f)$ is the one-sided power spectral density of the noise.
In our study, all simulations are performed in stationary Gaussian noise assuming Advanced LIGO \cite{TheLIGOScientific:2014jea} and Advanced Virgo \cite{TheVirgo:2014hva} design sensitivity  and we set $f_\textrm{low}$ and $f_\textrm{high}$ to $20$ Hz and $2048$ Hz.

The evidence $p(d|\mathcal{H})$ is given by
\begin{equation}
    p(d|\mathcal{H}) = \int_{\mathcal{V}} p(d|\vec{\theta},\mathcal{H}) p(\vec{\theta}|\mathcal{H}) d\vec{\theta},
\end{equation}
which is the normalization constant for the posterior distribution. The landscape of the likelihood function is explored using the nested sampling algorithm~\cite{Skilling:2006gxv} implemented in \textsc{parallel bilby}~\cite{Smith:2019ucc}.

\section{Waveform Approximants}\label{B}

Throughout this article, we used four different waveform approximants that are described in more detail below: 

\textit{TaylorF2 (TF2)} is a purely analytical model derived within the Post-Newtonian approximation, see Ref.~\cite{Blanchet:2013haa} for a review. 
The model includes point-particle~\cite{Sathyaprakash:1991mt,Blanchet:1995ez,Damour:2001bu,Blanchet:2004ek,Blanchet:2005tk,Mishra:2016whh} and aligned spin terms~\cite{Mikoczi:2005dn,Arun:2008kb,Bohe:2015ana,Mishra:2016whh} up to 3.5PN order as well as tidal effects up to 7.5PN order~\cite{Damour:2009wj,Vines:2011ud,Bini:2012gu,Damour:2012yf}.\footnote{We note that we employ the existing publicly available TF2 implementation in LALSuite that was used for the analysis of previous GW detections but does not incorporate recently updated 7PN and 7.5PN order terms~\cite{Henry:2020ski,Narikawa:2021pak}.}
 
\textit{IMRPhenomD\_NRTidal (PhenDNRT)} uses a binary black hole (BBH) baseline given by the phenomenological frequency-domain model IMRPhenomD, which was introduced in Refs.~\cite{Husa:2015iqa, Khan:2015jqa} and calibrated to untuned EOB waveforms~\cite{Taracchini:2013rva}, as well as numerical-relativity hybrids. 
The model describes spin-aligned systems throughout the inspiral, merger, and ringdown. 
The BBH model is augmented by the NRTidal phase description of Refs.~\cite{Dietrich:2017aum, Dietrich:2018uni}, which is based on analytical PN knowledge up to 6PN order, a tidal EOB model~\cite{Nagar:2018zoe, Bernuzzi:2014owa}, and numerical-relativity simulations. 
In PhenDNRT, no additional contributions from EOS-dependent spin-spin or cubic-in-spin effects are included. 

\textit{IMRPhenomD\_NRTidalv2 (PhenDNRTv2)} uses the same underlying BBH model as PhenDNRT, but has a different description of the tidal sector and the NRTidalv2 phase contribution is employed~\cite{Dietrich:2019kaq}. 
This tidal description incorporates PN information up to 7.5PN order and is calibrated to an updated set of tidal EOB and NR waveforms as outlined in Ref.~\cite{Dietrich:2019kaq}. 
These updates also include tidal contributions to the GW amplitude as well as the EOS-dependence on the quadrupole and octupole moment in the spin-spin (up to 3PN order) and cubic-in spin (up to 3.5PN order) contributions.

\textit{SEOBNRv4\_ROM\_NRTidalv2 (SEOBNRTv2)} also uses the NRTidalv2 description for the tidal sector which includes amplitude- and EOS-dependent spin effects. 
However, in contrast to PhenDNRTv2, the waveform model is based on an effective-one body description. 
The underlying BBH waveform model SEOBNRv4 was introduced in Ref.~\cite{Bohe:2016gbl} and its reduced-order model SEOBNRv4\_ROM is constructed following the methods outlined in Ref.~\cite{Purrer:2014fza}. 
The differences between PhenDNRTv2 and SEOBNRTv2 will enable us to understand the importance of the underlying BBH model for the inference of the EOS parameters.

\section{Assessment on systematic biases on mass and spin parameters}\label{app:ppplot}

\begin{figure*}[t]
    \centering
    \includegraphics[width=\textwidth]{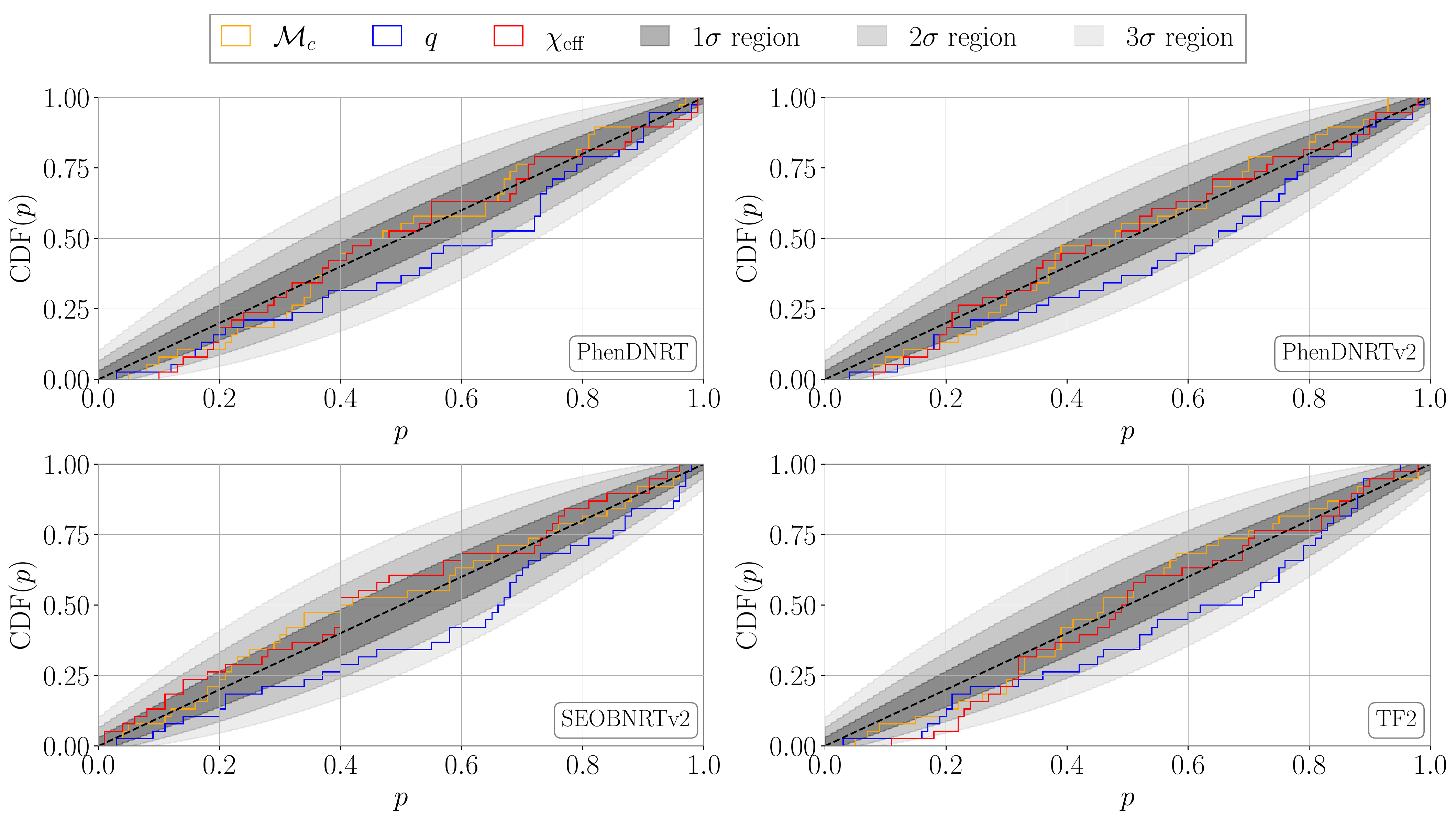}
    \caption{Percentile-percentile plot for chirp mass $\mathcal{M}_c$, mass ratio $q$, and $\chi_{\mathrm{eff}}$ for all waveform models. $p$ is the probability contained in a given credible interval and CDF($p$) is the fraction of the injections with their injected values laying inside that interval. The black dashed lines indicates the ideal 1-to-1 relation for the recovery of unbiased parameters. Shaded regions refer to $1$-, $2$-, and $3$-$\sigma$ regions.}
    \label{fig:pp-plot}
\end{figure*}

In order to investigate potential biases in the recovery of sources' mass and spin parameters, i.e., the chirp mass $\mathcal{M}_c$, the mass ratio $q$, and the effective spin parameter $\chi_{\mathrm{eff}}$, we used a percentile-percentile (pp) test. We expect our recovery to be unbiased when the pp-curve for each parameter follows the $x=y$ line, apart statistical fluctuations resulting from the Gaussian noise assumed in our BNS simulations.

We find that associated parameters analyzed in Fig.~\ref{fig:pp-plot} follow the expected $x=y$ relation, shown as black dashed line, and can be recovered within a 3-$\sigma$ region (2-$\sigma$ for PhenomDNRTv2). In order to quantify deviations seen in Fig.~\ref{fig:pp-plot}, we perform a Kolmogorov-Smirnov (KS) test for all models. Our KS statistic results, $D_n$, confirm that the deviations for the chirp mass, mass ratio, and the effective spin parameter are smallest for our injection model ranging up to $D_{n, \mathcal{M}_c} = 0.1$, $D_{n, q} = 0.17$, and $D_{n, \chi_{\mathrm{eff}}} = 0.1$, respectively.

As PhenomDNRT has the same point-particle description as PhenomDNRTv2, one expect it to have similar degree of performance as PhenomDNRTv2. Indeed, the PhenomDNRT's KS statistic are close to the one of PhenomDNRTv2. The KS statistic for chirp mass, mass ratio, and effective spin parameter are $D_{n, \mathcal{M}_c} = 0.11$, $D_{n, q} = 0.2$, and $D_{n, \chi_{\mathrm{eff}}} = 0.11$.

Because PhenomD reduces to TF2 at low frequency, the systematics induced by using TF2 should be less than the one by using SEOB. This matches our results with the TF2's KS statistic for chirp mass, mass ratio and effective spin parameter being $D_{n, \mathcal{M}_c} = 0.12$, $D_{n, q} = 0.19$, and $D_{n, \chi_{\mathrm{eff}}} = 0.12$, respectively. Because the tidal contribution is only included up to 7PN order for PhenomDNRT, while TF2 has included up to 7.5PN order, that could contributed to the marginally stronger systematics for mass ratio in PhenomDNRT.

While the largest deviation from the $x=y$ relation is present for SEOBNRTv2 ranging up to $0.12$ for $\mathcal{M}_c$, $0.22$ for $q$, and up to $0.17$ for $\chi_{\mathrm{eff}}$, we conclude that systematics are not pronounced for these source parameters regardless of waveform model in use.

\bibliography{main.bib}

\end{document}